\documentclass[aps,floatfix,twocolumn,showpacs,superscriptaddress,prl]{revtex4}
\usepackage{graphicx,bm}
\usepackage{amsmath}
\usepackage{amssymb}
\usepackage{hyperref}
\newcommand{\be}{\begin{equation}}
\newcommand{\ee}{\end{equation}}
\newcommand{\bea}{\begin{eqnarray}}
\newcommand{\eea}{\end{eqnarray}}
\newcommand{\rr}{\mathbf{r}}
\newcommand{\kk}{\mathbf{k}}

\newcommand{\apgt}{  \ {\raise-.5ex\hbox{$\buildrel>\over\sim$}}\ }
\newcommand{\aplt}{\ {\raise-.5ex\hbox{$\buildrel<\over\sim$}}\ }

\begin{document}
\title{Limit of Spin Squeezing in Finite Temperature Bose-Einstein Condensates}

\author{A. Sinatra}
\affiliation{Laboratoire Kastler Brossel, Ecole Normale Sup\'erieure, UPMC and CNRS,  Paris, France}
\author{E. Witkowska}
\affiliation{Institute of Physics, Polish Academy of Sciences, Warszawa, Poland }
\author{J.-C. Dornstetter}
\affiliation{Laboratoire Kastler Brossel, Ecole Normale Sup\'erieure, UPMC and CNRS,  Paris, France}
\author{Yun Li}
\author{Y. Castin}
\affiliation{Laboratoire Kastler Brossel, Ecole Normale Sup\'erieure, UPMC and CNRS,  Paris, France}

\begin{abstract}
We show that, at finite temperature, the maximum spin squeezing achievable using interactions in Bose-Einstein condensates has a finite limit when the atom number $N\to \infty$ at fixed density and interaction strength. We calculate the limit of the squeezing parameter for a spatially homogeneous system and show that it is bounded from above by the initial non-condensed fraction.  
\end{abstract}

\maketitle
Atomic clocks based on cold alkali atoms in two hyperfine states $a$ and $b$ are widely used as frequency standards. When  atoms in an uncorrelated quantum state are used, the
   clock precision is limited by the so-called projection noise, resulting from the quantum nature of the collective
   spin $\mathbf{S}$, {\sl i.e.}\  the sum of the effective spin 1/2 of each atom.
    This limit
   is actually already reached in most precise clocks \cite{Salomon}. Spin squeezing \cite{Ueda} amounts
   to creating quantum correlations among the atoms so as to increase the precision
   of the atomic clock beyond this standard quantum limit. The resulting gain for metrology is quantified 
 by the spin squeezing parameter $\xi^2$ \cite{Wineland}: The statistical uncertainty on the measured frequency when using a squeezed state is
$\Delta \omega_{\rm ab}=\xi \Delta \omega_{\rm ab}^{\rm unc}$, where
$\Delta \omega_{\rm ab}^{\rm unc}$ is the uncertainty using an uncorrelated state with the same atom number and Ramsey interrogation time. Spin squeezing in atomic ensembles was first obtained by quantum non-demolition measurements \cite{BigelowPolzik}.
Recently a significant amount of spin squeezing (up to 8 dB) has been achieved using atoms in a resonant optical
cavity \cite{Vuletic} or exploiting atomic interactions in bimodal Bose-Einstein condensates  
\cite{Oberthaler1,Oberthaler,Treutlein}. 
The ultimate limits of the different paths to spin squeezing are still an open question. We determine here the influence of the non-condensed fraction for spin squeezing schemes using condensates
\cite{Oberthaler,Treutlein,Sorensen_NaturePoulsen}. 

A central issue is the {\it scaling of the squeezing} for large atom numbers.
Most studies are based on a two-mode description \cite{Ueda}. In this case the squeezing parameter minimized over time $\xi^2_{\rm min}$ tends to zero (infinite metrology gain) for $N\to \infty$ as $\xi^2_{\rm min} \approx N^{-2/3}$.
The first analysis of squeezing at finite temperature \cite{Sorensen} used a large $N$
   expansion in a Bogoliubov-like  approach and could not predict any deviation
   of the spin squeezing scaling from the two-mode model. Here, using fully non-perturbative
   semi-classical field simulations and a powerful formulation of Bogoliubov
   theory in terms of the time dependent condensate phase operator \cite{supertruediff},
   we find on the contrary a dramatic effect of the multimode nature of the field:
   For a spatially homogeneous system in the thermodynamic limit, the two-mode
   scaling $\xi_{\rm min}^2\approx N^{-2/3}$ turns out to be completely irrelevant,
   and the spin squeezing has a finite optimal value that we determine analytically.
\begin{figure}[tb]
\centerline{\includegraphics[width=8.5cm,clip=]{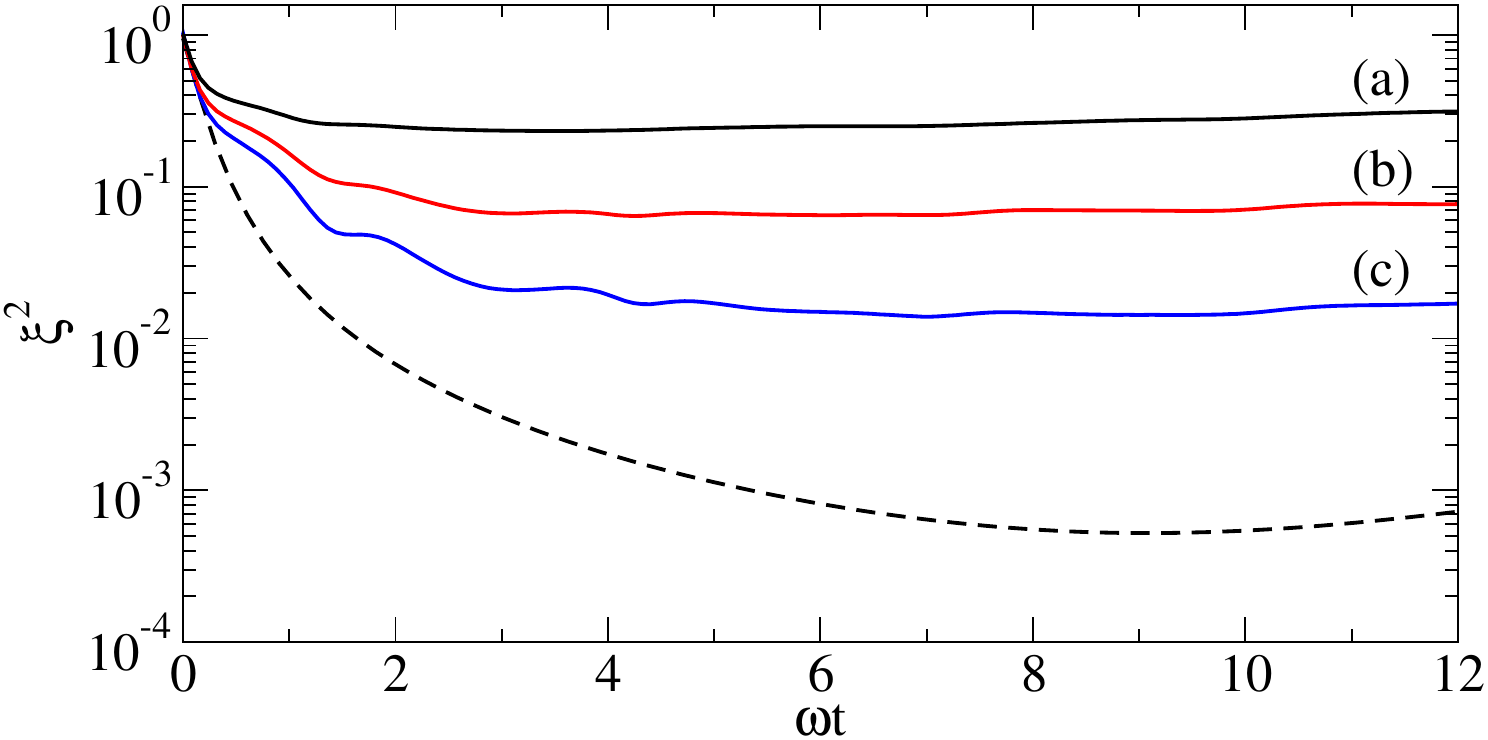}}
\caption{(Color online) Spin squeezing parameter $\xi^2$ as a function of time after the pulse mixing the states $a$ and $b$,
for  $N=10^{5}$ Rb atoms  ($s$-wave scattering length $a=5.3$\, nm), in a harmonic trap with oscillation frequency $\omega/2\pi=50$Hz.
The Thomas-Fermi chemical potential is  $\mu = 15.36 \hbar \omega$.
Finite temperature semi-classical field simulations  with initial non-condensed fractions:
(a) $\langle N_{\rm nc} \rangle/N=0.34$ (black solid line), (b) $\langle N_{\rm nc} \rangle/N=0.20$ (red line), 
(c) $\langle N_{\rm nc} \rangle/N=0.09$ 
(blue line), corresponding to $k_B T/\mu=2.08,1.53,1.17$ respectively. Dashed line: Two-mode theory for comparison.
\vspace{-0.25cm}
\label{fig:trap}}
\end{figure}

The physical problem that we face is the dynamical evolution of a finite temperature Bose-condensed gas
after a $\pi/2$ pulse that puts each atom in a coherent superposition of two internal states $a$ and $b$.
This produces a non-equilibrium state that has
a non-trivial evolution due to the atomic interactions inside each internal state.
For simplicity, it is assumed that there is no cross-interaction between $a$ and $b$ atoms \cite{g_ab=0}.

{\it Semi-classical field simulations - }
In Fig.\ref{fig:trap} we compare the two-mode theory with semi-classical field simulations at finite temperature in a trap. 
The gas is initially in state $a$ at thermal equilibrium at a temperature (times $k_B$) larger than its chemical potential. 
As we expect the physics to be dominated by low energy excitations, where thermal
 fluctuations dominate over quantum fluctuations, we use for state $a$
 a classical field description \cite{KaganSachdev,Rzazewski0Burnett,Svistunov} with an energy cut-off at $k_B T$.
The initial field $\psi_a^{(0)}$ then randomly samples the thermal equilibrium
 classical field distribution for the canonical ensemble at temperature $T$.
 For the initially empty state $b$,  inspired by the truncated Wigner approach
\cite{DrummondWigner} we represent the vacuum by a classical field $\psi_b^{(0)}$
 having  in each mode independent Gaussian complex fluctuations of zero mean and variance $1/2$.
 A sudden $\pi/2$ pulse mixes the initial fields $\psi_a^{(0)}$ and
 $\psi_b^{(0)}$ so that, at time $t=0^+$, just after the pulse:
\begin{equation}
{\psi}_{a,b}(0^+)= [ {\psi}_{a,b}^{(0)} \mp {\psi}_{b,a}^{(0)} ] / \sqrt{2} \label{eq:pulse} \,.
\end{equation}
 At later times, the fields, that evolve independently according to the non-linear Schr\"odinger equation ($\nu=a,b$)
\begin{equation}
i \hbar \, \partial_t \psi_{\nu}= \left[ -\frac{\hbar^2 \Delta}{2m} + \frac{1}{2} m \omega^2 {\bf r}^2
+ g |\psi_{\nu}({\bf r},t)|^2  \right] \, \psi_{\nu} \, 
\label{eq:sc}
\end{equation}
are used to calculate the spin components (\ref{eq:Sxyz}).
Fig.\ref{fig:trap} corresponds to a harmonically trapped gas with same oscillation frequency
 $\omega$ and same coupling constant $g=4\pi\hbar^2 a/m$ for the two internal states,
 where $a$ is the $s$-wave scattering length.
The squeezing is created dynamically by the interactions. Still, even for a moderate
non-condensed fraction $\langle N_{\rm nc} \rangle/N=0.09$, $\xi_{\rm min}^2$ exceeds by a factor 
$\simeq 30$ the two-mode value. 

In order to isolate the effect of the non-condensed fraction from spatial dynamics effects taking place in the trapped system \cite{Spsqdyn},
and to develop an analytical theory, we consider from now-on the homogeneous case. We first use a semi-classical field model 
that can be simulated exactly, and we generalize the results to the quantum field in the end.
The real space is discretized on a lattice with unit cell of volume $dV$, within a volume $V$ with periodic boundary conditions \cite{supertruediff}.
The Hamiltonian after the pulse for component $a$ (and similarly for $b$) reads:
\begin{equation}
H_a={ \sum_{\kk \in {\rm BZ}}} \frac{\hbar^2 k^2}{2m} a^\ast_\kk a_\kk + 
		\frac{g}{2} dV { \sum_{\rr}}  
	\left| \psi_a(\rr) \right|^4 \,,
\label{sinatra_eq:discrHam}
\end{equation}
where ${\rm BZ}$ is the first Brillouin zone of the lattice and the
largest kinetic energy in ${\rm BZ}$ is $k_BT$.
The fields have Poisson brackets
$i\hbar \{\psi_\mu(\rr),\psi_\nu^*(\rr')\}= \delta_{\rr \rr'} \delta_{\mu \nu}/dV$
 with $\mu,\nu=a$ or $b$, and
${a}_\kk ({b}_\kk)$ is the amplitude of ${\psi}_{a,b}$ over the plane wave of momentum $\kk$. 
In terms of the fields, the collective spin components are 
\begin{equation}
S_x+i S_y=dV {\textstyle \sum_{\rr}}  \,  {\psi}^\ast_a
(\rr\,){\psi}_b(\rr\,), \hspace{0.5cm} S_z=\frac{N_a-N_b}{2} \label{eq:Sxyz}
\end{equation}
with $N_\nu=dV \sum_\rr |\psi_\nu(\rr)|^2,$ $\nu=a,b$.
The spin squeezing parameter $\xi^2$ is defined as
\begin{eqnarray}
\xi^2(t)&=& { \Delta S_{\perp,{\rm min}}^2(t) \over \langle S_x(t)\rangle^2 } \times
                   { \langle S_x(0^+)\rangle^2  \over \Delta S_{\perp,{\rm min}}^2(0^+)}  \,, \label{eq:spsqdef}\\
\Delta S_{\perp,{\rm min}}^2 &=& {1\over 2}\left[ \langle S_y^2 \rangle + \langle S_z^2 \rangle - |\langle (S_y + i S_z)^2 \rangle| \right]\,,
\end{eqnarray}
where $\Delta S_{\perp,{\rm min}}^2$ is the minimal variance of the spin orthogonally to its mean direction and
$\langle S_x\rangle$ is the mean spin length. 
The ratio ${ \Delta S_{\perp,{\rm min}}^2(t) / \langle S_x(t) \rangle^2 }$ is proportional
to the statistical variance $\Delta \omega_{ab}^2$ of the measured frequency if the system state at time $t$ is used as an input in
an atomic clock, after an appropriate rotation, see Eq.(22) in \cite{Wineland}. At $t=0^+$, for the initially uncorrelated spins
this ratio is equal to $(\Delta \omega_{ab}^{\rm unc})^2$.

\begin{figure}[tb]
\centerline{\includegraphics[width=8.5cm,clip=]{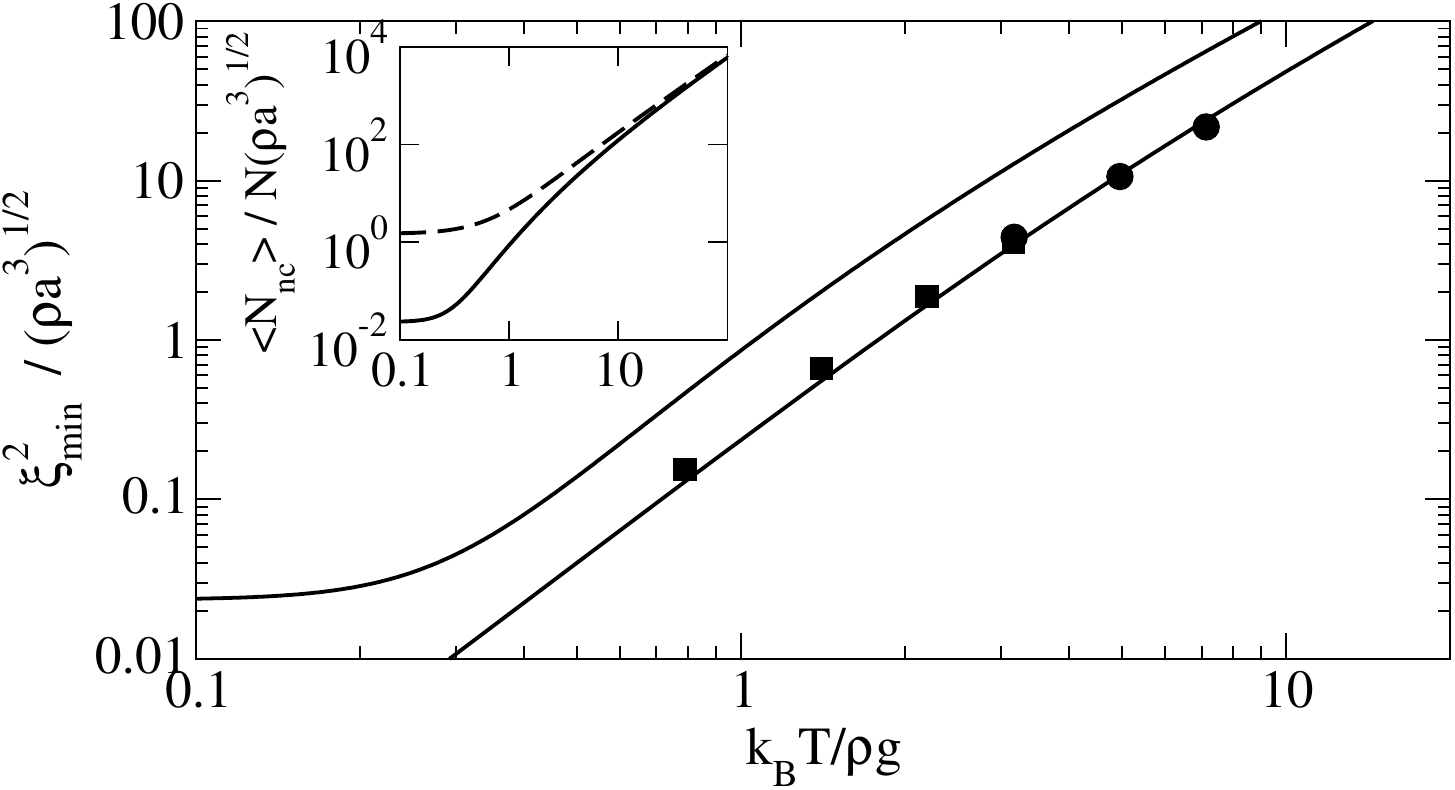}}
\caption{Best squeezing $\xi_{\rm min}^2$ divided by $\sqrt{\rho a^3}$ as a function of $k_BT/\rho g$.
Symbols: Semi-classical field simulations with $\sqrt{\rho a^3}=1.32 \times10^{-2} $ (squares) and $1.94 \times10^{-3}$
(disks) in the thermodynamic limit (reached already for $N=3\times 10^{4}$ except for the lowest value of $k_BT/\rho g$). Lower solid line: 
Analytical semi-classical field result (\ref{eq:xibest}). Upper solid line: Quantum result (\ref{eq:xibest_q}). Inset: 
Quantum $\xi_{\rm min}^2$ (solid line) and non-condensed fraction $\langle N_{\rm nc} \rangle/N$ (dashed line), both divided by $\sqrt{\rho a^3}$, as functions of $k_BT/\rho g$.
\label{fig:xibest}}
\end{figure}

We first performed semi-classical field simulations for different temperatures and increasing system sizes \cite{genuine}.
The result (not shown) is that $\xi_{\rm min}^2$ converges to a finite value 
in the thermodynamic limit: $N$$\to$$\infty$, $\rho,g,T$=constant, where $\rho=N/V$ is the total density.
Five independent physical parameters are in the model,
$ \hbar^2/m, g, k_BT, N$ and $ V$. 
From dimensional analysis, $\xi_{\rm min}^2$ is thus
a function of the three independent dimensionless quantities
$N, \sqrt{\rho a^3},$ and  $k_BT/\rho g$. The existence of
a thermodynamic limit then implies $\xi_{\rm min}^2=f\left(\sqrt{\rho a^3}, {k_BT/\rho g}\right)$.
Second, we performed simulations increasing the density in the weakly interacting limit \cite{CastinDum},
$\rho \to \infty$, $g\to 0$ with $T,\rho g$=constant. The result (not shown) is that $\xi_{\rm min}^2$ scales as $1/\rho \propto \sqrt{\rho a^3}$. This implies:
\begin{equation}
\xi_{\rm min}^2/\sqrt{\rho a^3}=F(k_BT/\rho g) \,. \label{eq:universal}
\end{equation}
In Fig.\ref{fig:xibest} we show the universal behavior (\ref{eq:universal}). Disks and squares correspond to two different values
of $\sqrt{\rho a^3}$ in simulations. 

{\it Semi-classical field analytics -}
We now develop an analytical theory to explain these results. 
We split the fields after the pulse as ${\psi_a} = \frac{{a}_{\bf 0}}{\sqrt{V}} + {\psi}_{a \perp}$ and similarly for $\psi_b$.
We introduce the modulus and phase conjugate variables 
\begin{equation}
{a}_{\bf 0}=e^{i {\theta}_a} \sqrt{ N_{a{\bf 0}} } \:\: \:\: ,  \:\:\:\: {b}_{\bf 0}=e^{i {\theta}_b} \sqrt{ N_{b{\bf 0}} }\, \,,
\end{equation}
for the condensate modes, and number conserving non-condensed fields $\Lambda_a$ and $\Lambda_b$ \cite{CastinDum} that we expand over Bogoliubov
modes with amplitudes $c_{a\kk}$ and $c_{b\kk}$ \cite{note_Bogfix}:
\begin{eqnarray}
\Lambda_a=e^{-i\theta_a}\psi_{a \perp} = {\textstyle \sum_{\kk \in {\rm BZ}}^{\kk \neq {\bf 0}}}  \left( U_k c_{a\kk}+V_k c_{a-\kk}^\ast \right) \frac{e^{i\kk \cdot \rr}}{\sqrt{V}} &&\\
U_k+V_k=\left( {E_k \over E_k + \rho g}\right)^{1/4} ; \hspace{0.5cm} E_k={\hbar^2k^2 \over 2m} \,. && \label{eq:sk}
\end{eqnarray}
The spin component $S_y$ is given by
\begin{equation}
S_y={\rm Im} \left[ e^{-i\left( \theta_a-\theta_b \right)} \left(\sqrt{N_{a{\bf 0}}N_{b{\bf 0}}} + dV {\textstyle \sum_{\rr}} \, \Lambda_a^\ast \Lambda_b \right) \right] .
\label{eq:Sy}
\end{equation}
Our strategy is to perform a {\it double expansion} of $\langle S_y^2 \rangle$ and $\langle S_y S_z \rangle$. We will need terms up to $\sim N$
in the thermodynamic limit and up to order one in the non-condensed fraction 
$\langle N_{\rm nc} \rangle/N$. At this order, we can replace $\langle S_x(t) \rangle$ in the denominator of
(\ref{eq:spsqdef}) by its value at $t=0^+$ so that $\xi^2(t) \simeq (4/N) \Delta S^2_{\perp,{\rm min}}(t)$.
In the Bogoliubov approximation, the condensate phases at $t>0$ obey \cite{supertruediff}
\begin{eqnarray}
\theta_a-\theta_b&=& - 2 {{\rm Im \,}b_{\bf 0}^{(0)} \over \sqrt{N_{a_{\bf 0}}^{(0)}}} - \frac{gt}{\hbar V} 
\left[ ( N_a - N_b ) + {\cal D} \right]\label{eq:phase} \\
{\cal D} &=& {\textstyle \sum_{\kk \in {\rm BZ}}^{\kk \neq {\bf 0}}} \left( U_k + V_k \right)^2 \left( |c_{a \kk}|^2 - |c_{b \kk}|^2\right) \,. \label{eq:calS}
\end{eqnarray}
The {\it multimode part}  ${\cal D}$ of the relative phase derivative is absent in the two-mode theory.
In thermodynamic limit $\theta_a-\theta_b \sim 1/\sqrt{N}$ and it is sufficient to expand the exponential in
(\ref{eq:Sy}) to second order. Introducing the constant of motion $N_{\rm tot}=N + \sum_{{\bf k}\in {\rm BZ}} |b_{\bf k}^{(0)}|^2$,
we also expand:
\begin{equation}
\sqrt{N_{a{\bf 0}}N_{b{\bf 0}}}\simeq {N_{\rm tot} \over 2} - {1 \over 2} dV {\textstyle \sum_{\rr}} \, \left( |\Lambda_a|^2 + |\Lambda_b|^2\right)\,.
\end{equation}

{\it Best squeezing -}
We note that $S_y$ has a leading contribution that 
grows linearly in time while $S_z$ is constant.
One finds that $\xi^2$ reaches its minimal value at long times where $|S_y|\gg |S_z|$ and to leading (zeroth) order in $1/t$:
\begin{equation}
\xi^2(t) \simeq \frac{\langle S_y^2\rangle \langle S_z^2 \rangle - \langle S_y S_z \rangle^2}
{\langle S_y^2 \rangle \langle S_z^2 \rangle} \simeq {\langle {\cal D}^2 \rangle \over N} =\xi_{\rm min}^2 \,. \label{eq:xi2S}
\end{equation}
Remarkably, the best squeezing (\ref{eq:xi2S}) only involves the multimode part (\ref{eq:calS}) of the phase difference. 
Let us sketch the physics. Due to interactions, the relevant contribution to $S_y$ coming from the phase difference (\ref{eq:phase}), 
becomes proportional to $(2S_z+{\cal D})t$ at long times.  If the multimode contribution ${\cal D}$ was absent, this would allow a perfect cancellation between the correlation $\langle S_y S_z \rangle^2$ and the
product $\langle S_y^2 \rangle \langle S_z^2 \rangle$ in (\ref{eq:xi2S}) leading to $\xi_{\rm min}^2=0$ in the thermodynamic limit. In presence of ${\cal D}$, this is not possible: Interactions with the excited modes affect the relative phase, thus $S_y$ \cite{crossed_term}.
An explicit calculation gives:
\begin{equation}
\xi_{\rm min}^2 =   
\int_{\rm BZ}\! {d^3 k \over (2\pi)^3} \, \frac{s_k^4}{2\rho} \, n_k^{(0)} \left(  {(s_k^{(0)} )^2 \over  s_k^4 }  + {s_k^4 \over (s_k^{(0)})^2 }  \right)
\label{eq:xibest}
\end{equation}
(lower line in Fig.\ref{fig:xibest}). $s_k=U_k+V_k$, $s_k^{(0)}$ is the same quantity before the pulse obtained by replacing $\rho g$ with $2 \rho g$ in (\ref{eq:sk}); $n_k^{(0)}=k_BT/\epsilon_k^{(0)}$ 
with $\epsilon_k^{(0)}=[E_k(E_k+2\rho g)]^{1/2}$. 

\begin{figure}[tb]
\centerline{\includegraphics[width=8.5cm,clip=]{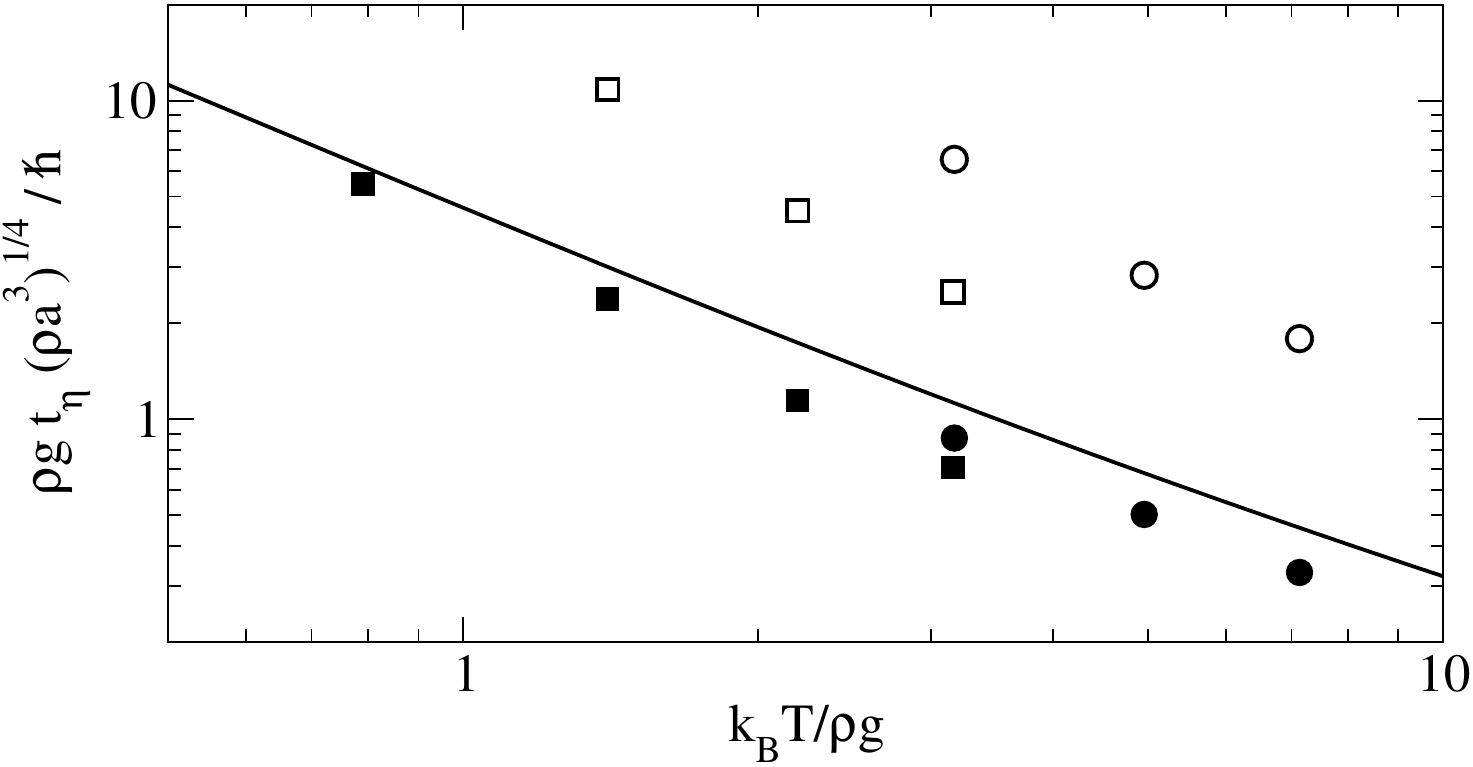}}
\caption{``Close to best" squeezing time $t_\eta$ for $\eta=0.2$ from simulations (filled symbols) and thermalization time $t_{\rm therm}$ extracted  from the decay of the contrast $\langle S_x \rangle$ in simulations (open symbols). $\sqrt{\rho a^3}=1.32 \times10^{-2} $ (squares) and $1.94 \times10^{-3}$ (disks/circles). Line: Analytical result (\ref{eq:t_eta}) using (\ref{eq:xibest}). 
\vspace{-0.5cm}
\label{fig:tratio}}
\end{figure}

{\it Squeezing time -}
From the next to leading order in the time expansion of the squeezing parameter the best squeezing is reached in an infinite time:
\begin{equation}
\xi^2(t)=\xi_{\rm min}^2  + \left({\hbar\over \rho gt}\right)^2\left[ 1 + O\left( {\langle N_{\rm nc}\rangle \over N} \right)\right] +O\left( t^{-4}\right). \label{eq:xi(t)}
\end{equation}
This is a limitation of the analytical approach. Still, since the numerical squeezing
  curve $\xi^2(t)$ is quite flat around its minimum, it
  suffices to determine a ``close to best" squeezing time $t_\eta$ defined
  as $\xi^2(t_\eta)=(1+\eta) \xi_{\rm min}^2$, where $\eta>0$. Then, according to (\ref{eq:xi(t)}),
  $t_\eta$ is finite and given by
\begin{equation}
\frac{\rho g}{\hbar} t_\eta = {1 \over \sqrt{\eta \xi^2_{\rm min}}} \,.\label{eq:t_eta}
\end{equation}
The ``close to best" squeezing time $t_\eta$ (\ref{eq:t_eta}) for $\eta=0.2$ is shown in Fig.\ref{fig:tratio} and compared to simulations.

\begin{figure}[tb]
\centerline{\includegraphics[width=8.5cm,clip=]{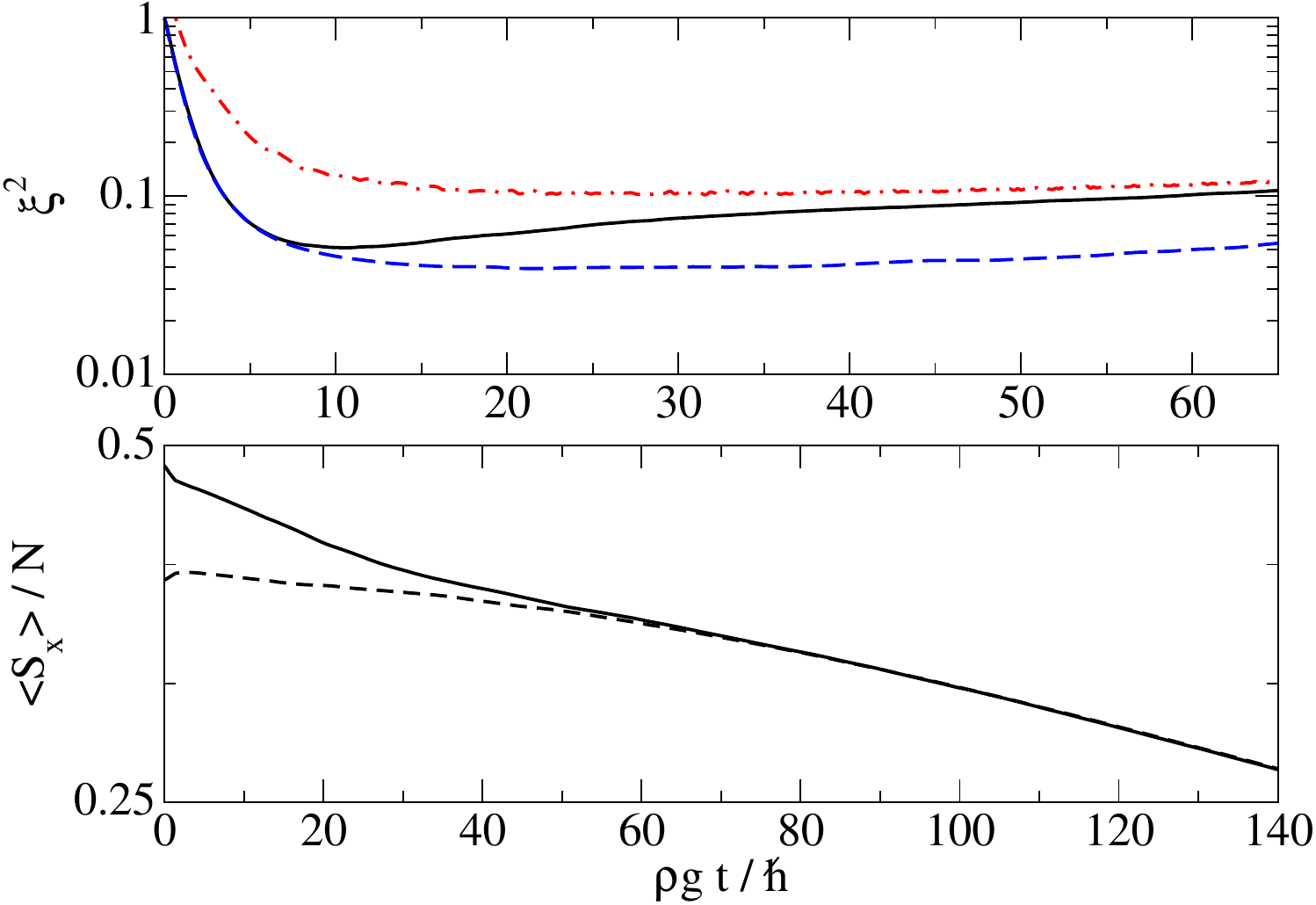}}
\caption{(Color online) Top: Spin squeezing $\xi^2$  as a function of time. Black solid line: Simulation.
Blue dashed line: Bogoliubov theory. Red dash-dotted line: Ergodic model.
Bottom: Total contrast $\langle S_x \rangle$ (solid line) and condensate contrast ${\rm Re} \langle b_{\bf 0}^\ast a_{\bf  0} \rangle$ 
(dashed line) as functions of time. 
$N=3\times10^4$, $k_BT/\rho g=3.16$, $\sqrt{\rho a^3}=1.32\times 10^{-2}.$
\vspace{-0.25cm}
\label{fig:therm}}
\end{figure}

A last issue is that of {\it thermalization} that brings the system back to equilibrium after the pulse,
neglected in Bogoliubov theory and in our analytics
but included in the semi-classical field simulations.
It is thus possible to reach $\xi^2 = (1+\eta) \xi_{\rm min}^2$ with $\xi_{\rm min}^2$ given by (\ref{eq:xibest}) only if
$t_\eta$ given by (\ref{eq:t_eta}) is shorter than the thermalization time:
\begin{equation}
t_\eta < t_{\rm therm} \,. \label{eq:cond_t}
\end{equation}
We first estimate $t_{\rm therm}$ from Landau-Beliaev damping rates of modes at energies $k_BT$ or $\rho g$.
For fixed $k_BT/\rho g$, this gives $\rho g t_{\rm therm}/\hbar \propto (\rho a^3)^{1/2}$ \cite{genuine,Giorgini}. Consequently 
$t_\eta/t_{\rm therm}$ scales as $(\rho a^3)^{1/4}$ and (\ref{eq:cond_t}) is satisfied in the weakly interacting limit.
In Fig.\ref{fig:therm} we show the squeezing parameter $\xi^2$ and contrast $\langle S_x \rangle$ 
during thermalization.
For $\xi^2(t)$, we compare the simulation with (i) the full Bogoliubov theory (without the analytic expansions) that we implement numerically for a finite size system and (ii) a Bogoliubov ergodic model \cite{supertruediff} where the amplitudes $c_{a\kk}$, $c_{b\kk}$ in (\ref{eq:phase}) sample microcanonical distributions with number of particles and energy $(N_a,E_a)$, $(N_b,E_b)$ that are constant after the pulse.
The simulation agrees with the Bogoliubov model at short times (including the ``close to best"  squeezing time) and then converges
towards the ergodic model. In Fig.\ref{fig:tratio} we show the thermalization time $t_{\rm therm}$ obtained from simulations (empty symbols). 
It scales with $\rho a^3$ as expected from the Landau-Beliaev estimate 
and is $\gg  t_{\eta}$ which justifies our analytical approach.
We extract $t_{\rm therm}$ as the relaxation time of the total contrast $\langle S_x \rangle$ towards the condensate contrast
${\rm Re} \, \langle b_{\bf 0}^\ast a_{\bf 0} \rangle$, see Fig.\ref{fig:therm} (bottom):
As thermalization occurs, the excited modes dephase and
\begin{equation}
\langle S_x \rangle = {\textstyle \sum_{{\kk}\in{\rm BZ}} }\;  {\rm Re} \,\langle b_{\kk}^\ast a_{\kk} \rangle \underset{t > t_{\rm therm}}{\simeq} 
{\rm Re}  \, \langle b_{\bf 0}^\ast a_{\bf 0} \rangle\,.
\end{equation} 
The longer time scale for the decay of $\langle b_{\bf 0}^\ast a_{\bf 0} \rangle$ 
is set by partition noise phase spreading \cite{QPS} plus thermal effects.
{\it Quantum field - }
Our analytic calculations for the semi-classical field can be generalized to the quantum field: Relations $\xi_{\rm min}^2=
\langle {\cal D}^2 \rangle/N$ and (\ref{eq:t_eta}) are unchanged and
\begin{equation}
\xi_{\rm min}^2 \!=\!\!
\int\!\! {d^3 k \over (2\pi)^3} \, {s_k^4 \over 2\rho} \left[ \Big(n_k^{(0)}+{1\over 2}\Big)\! 
\left( {  (s_k^{(0)})^2\over  s_k^4 } + {s_k^4 \over  (s_k^{(0)})^2 }  \right) -1 \right] \,,
\label{eq:xibest_q}
\end{equation}
$n_k^{(0)}$ are Bose occupation numbers of Bogoliubov modes in internal state $a$ before the pulse. 
At zero temperature:
\begin{equation}
{\xi_{\rm min}^{2 \, (T=0)} \over \sqrt{\rho a^3} }= \sqrt{{8\over \pi}} \left[ {19\over 6}\sqrt{2} -{3\over 2} {\rm ln}(\sqrt{2}+1)-\pi\right] \simeq 0.02344 \label{eq:xi2T0} \,.
\end{equation}
In practice  $\rho a^3 < 10^{-6}$ in present squeezing experiments so that (\ref{eq:xi2T0}) predicts $\xi_{\rm min}^{2 \, (T=0)}  \aplt 2. 10^{-5}$.
This value is very low, in particular below the limit given by particle losses \cite{PRLlosses}.
Interestingly, $\xi^2_{\rm min}$ is lower than the initial non-condensed fraction $\langle N_{\rm nc} \rangle/N$ at any temperature, see the inset of Fig.\ref{fig:xibest}, and $\xi_{\rm min}^{2} \sim \langle N_{\rm nc} \rangle/N$ for $k_BT \gg \rho g$.  
Already for $k_BT/\rho g=2$, $\xi_{\rm min}^2$ and $\langle N_{\rm nc}\rangle/N$ are within a factor three. 
A similar conclusion seems to hold in a trap, see Fig.\ref{fig:trap}. The recent experiments \cite{Oberthaler,Treutlein},
where the non-condensed fraction is expected to be lower than the measured  $\xi^2$,  
were probably limited by losses for \cite{Oberthaler} and technical noise for \cite{Treutlein},
rather than by finite temperature.

In conclusion, contrarily to the prediction of the two-mode model, the best spin squeezing $\xi_{\rm min}^2$ admits a finite limit for $N\to \infty$
at fixed density and interaction strength, that we determine analytically. 
Physically this is due to the perturbation of the condensate phase evolution by interactions with the thermally populated modes.

We thank J.Est\`eve for discussions and acknowledge support from CNRS and Polish GRF: N N202 104136, 2009-2011. J.-C.D. is a member of 
Ecole Polytechnique.


\begin{thebibliography}{99}
\bibitem{g_ab=0} This may be done by spatial separation of the spin states \cite{Treutlein}
or by Feshbach tuning of the $a$-$b$ scattering length \cite{Oberthaler}.
\bibitem{note_Bogfix} Replacing $N_a,N_b$ by $N/2$ in the Bogoliubov modes is accurate for the timescales we are interested in.
\bibitem{crossed_term} Due to the determinant structure of the numerator 
$\langle S_y^2\rangle \langle S_z^2 \rangle - \langle S_y S_z \rangle^2$ in (\ref{eq:xi2S}), 
the crossed term $\langle S_z {\cal D}\rangle$ appears squared and is negligible at the present order.
\end{thebibliography}
\end{document}